\documentclass[prc,showpacs,showkeys,nofootinbib,superscriptaddress]{revtex4}
\usepackage{graphicx,amssymb,color}

\begin{document}

\title{Low-energy neutrinos at off-axis from a standard beta-beam}
\author{R.~Lazauskas}
\email{lazauska@lpsc.in2p3.fr}
\affiliation{Institut de Physique Nucl\'eaire, F-91406 Orsay cedex, France}
\author{A.~B.~Balantekin}
\email{baha@physics.wisc.edu}
\affiliation{Department of Physics, University of Wisconsin,
Madison, WI 53706, USA}
\author{J.~H.~de~Jesus}
\email{jhjesus@physics.wisc.edu}
\affiliation{Department of Physics, University of Wisconsin,
Madison, WI 53706, USA}
\author{C.~Volpe}
\email{volpe@ipno.in2p3.fr}
\affiliation{Institut de Physique Nucl\'eaire, F-91406 Orsay cedex, France}
\date{\today }

\begin{abstract}
We discuss a scenario to extract up to 150~MeV neutrinos at a standard
beta-beam facility using one and two detectors off-axis. In particular we show that the high-energy component of the
neutrino fluxes can be subtracted through a specific combination
of the response of two off-axis detectors. A systematic analysis
of the neutrino fluxes using different detector geometries is
presented, as well as a comparison with the expected fluxes at a
low-energy beta-beam facility. The presented option
could offer an alternative way to perform low-energy neutrino
experiments.
\end{abstract}

\maketitle

\section{Introduction}
\label{sec:introduction}
Low-energy neutrino sources play an important role in neutrino physics.
Depending on the applications, one typically can choose between
single-spectrum but intense, sources such as reactors, or multi-spectra
sources with lower intensity such as the proposed low-energy beta-beam
facilities. In such facilities one can study neutrino-nucleus interactions,
fundamental neutrino properties, and perform various electroweak tests.

Low-energy neutrino-nucleus interactions are important in several
contexts. The response of the chemical detectors to low-energy
neutrino sources such as the Sun and supernovae is dependent on
their neutrino capture cross sections.  Neutrino-nucleus cross
sections are an important ingredient in understanding various
astrophysical phenomena, for example, the dynamics of the
core-collapse supernovae~\cite{Balantekin:2003ip}, calculating the
yields of the supernova r-process
nucleosynthesis~\cite{Meyer:1998sn}, and assessing the formation
possibility of a black hole from the fossil abundances of the
r-process elements~\cite{Sasaqui:2005rh}.  These interactions are
also an important input into models of gamma-ray
bursts~\cite{Ruffert:1996by,Kneller:2004jr} and their
understanding is fundamental to the observation of neutrino
signals from astrophysical
sources~\cite{Vogel:1999zy,Beacom:2002hs}.  In particle physics,
low-energy neutrinos can be used as probes to test the electroweak
component of the Standard Model~\cite{'tHooft:1971ht}.

Low-energy beta-beam facilities first proposed in Ref.~\cite{Volpe:2003fi} yield pure beams of electron neutrinos or antineutrinos produced through the decay of radioactive ions circulating
in a storage ring~\cite{Zucchelli:2002sa, Volpe:2006in}. Several
applications utilizing low-energy beta-beams have been discussed in the
literature concerning neutrino-nucleus
scattering~\cite{Serreau:2004kx,McLaughlin:2004va,Volpe:2005iy},
electroweak tests of the Standard Model
\cite{McLaughlin:2003yg,Balantekin:2005md,Balantekin:2006ga,Bueno:2006yq,Barranco:2007tz},
as well as core-collapse supernova
physics~\cite{Volpe:2003fi,Jachowicz:2005ym}.

An extensive analysis of the physics potential of a beta-beam
facility is currently ongoing, in parallel with the feasibility
design study. One of the primary goals of a standard beta-beam
facility is to test CP-violation in the lepton sector.  Currently, experiments are in the planning stage to measure the third mixing angle, $\theta_{13}$, at reactors~\cite{Ardellier:2006mn,unknown:2007ug}.  Beta-beam facilities will be able to measure this angle as well as the associated CP-violating Dirac phase.  Various scenarios for beta-beam facilities have been considered as far as the measurement of $\theta_{13}$ and CP violation in the lepton sector is concerned~\cite{Mezzetto:2003ub, Mezzetto:2005yf, Mezzetto:2005ae, Burguet-Castell:2003vv, Burguet-Castell:2005pa, Campagne:2006yx, Agarwalla:2005we, Agarwalla:2006gz, Agarwalla:2006vf, Donini:2005qg, Donini:2004hu, Donini:2004iv, Donini:2003vz, Donini:2002rm, Huber:2005jk}.  Lepton number violating interactions are discussed in Ref.~\cite{Agarwalla:2006ht} (for a review, see Ref.~\cite{Volpe:2006in}).

In this paper we discuss an alternative to using low-energy beta-beams,
namely the possibility of extracting low-energy neutrino fluxes emitted
from a standard beta-beam facility, using a specific off-axis configuration.
The basic idea here is that if a detector is placed away from the
principal axis of the storage ring, it will detect the least energetic neutrinos
emitted from the parent nucleus. We also explore possible geometries for
such off-axis detectors. From the practical point of view, the proposed
idea presents several attractive features. First it would not require a devoted storage ring
as discussed in previous works
\cite{Volpe:2003fi,Serreau:2004kx,McLaughlin:2004va,Balantekin:2005md,Balantekin:2006ga,Bueno:2006yq,Barranco:2007tz,Jachowicz:2005ym}.
Second, specific beamtime would not be needed since one would exploit the
neutrino beams planned for CP violation studies. Third, one would benefit
from their good duty cycle that would
help reducing the atmospheric background.

The main body of this paper is Section~\ref{sec:profiles}. In this section
we derive the formulas and discuss the neutrino flux profiles for three
different scenarios: the original low-energy beta-beam scenario, used as reference (Section~\ref{sec:referencescenario}); one off-axis detector in a standard
 beta-beam facility (Section~\ref{sec:offaxis}); and two off-axis
detectors in a standard beta-beam facility (Section~\ref{sec:twodetectors}).
A discussion of our results is then given in
Section~\ref{sec:conclusions}.

\section{Neutrino flux profiles}
\label{sec:profiles}
Zucchelli first proposed the idea to use boosted radioactive ions as a new
method to produce pure, collimated and well known electron (anti)neutrino
fluxes~\cite{Zucchelli:2002sa}. The ions are stored in a storage ring where
they decay. To get the fluxes one needs to integrate over the storage ring
straight sections and the volume of the detector. The average neutrino flux
at the detector is therefore given by (the precise formalism can be
found in \cite{Serreau:2004kx} and also in \cite{Amanik:2007zy})
\begin{equation}
\label{eqn:flux}
\widetilde{\Psi }_{tot}(E_{\nu }) = f \tau \int_{0}^{Z}
\frac{dl}{L}\int_{V}\frac{\Psi _{lab}(E_{\nu },\hat{r} )}{4\pi r^{2}}\,dV~,
\end{equation}
where $L$ is the total and $Z$ is the straight section lengths of the
storage ring. In the stationary regime, the mean number of ions in the
storage ring is $\gamma \tau f$, where $\tau= t_{1/2} /\ln 2$ is the
lifetime of the parent nucleus and $f$ is the number of injected ions per
unit time. In Eq.~(\ref{eqn:flux}), the integration is performed over
the detector volume $V$ and the nearest storage ring straight section,
with $\vec{r}$ being the vector connecting two points in the storage ring
and the detector.

Using Eq.~(\ref{eqn:flux}), the total number of events per unit time with
energies up to $E_{\rm max}$, in the detector, is
\begin{equation}
\label{eqn:dndt}
dN/dt =n \int_{0}^{E_{\max }}\widetilde{\Psi }_{tot}(E_{\nu })
\sigma(E_{\nu})dE_{\nu }~,
\end{equation}
with $n$ being the number of target particles per unit volume and
$\sigma(E_{\nu})$ the cross section.

The calculations we present are performed assuming parameters from
the currently ongoing feasibility
study~\cite{Autin:2002ms,betabeam}.  For the original beta-beam
scenario, we assume a storage ring of total length $L=6580$~m with
straight sections of length $Z=2501$~m. The $^6$He expected
intensity to produce antineutrino fluxes is $f=2.53\times
10^{13}$~ions/s.  For the low-energy beta-beam, we consider
$L=1885$~m and $Z=678$~m with $f=2.65\times
10^{12}$~ions/s~\cite{chancepayet}. 
It is important to emphasize that these ion intensities come from the 
first feasibility study and are very preliminary. In particular, a new production
method has been proposed recently which might give increased intensities
\cite{Rubbia:2006pi}.

\subsection{Reference scenario: the low-energy beta-beam facility}
\label{sec:referencescenario}
The idea of a low-energy beta-beam
facility producing neutrinos in the 100~MeV energy range has been first proposed in~\cite{Volpe:2003fi}.  The disposal of a devoted storage ring would
probably be the ideal tool for low-energy neutrino
physics~\cite{Serreau:2004kx}.  The potential of such a facility
has been stressed in several papers~
\cite{McLaughlin:2004va,Volpe:2005iy,Balantekin:2005md,Balantekin:2006ga,Bueno:2006yq,Volpe:2003fi,Barranco:2007tz,Jachowicz:2005ym}.
  In Table~\ref{tbl:lowenergy}, we
summarize the characteristics of the corresponding fluxes for the
case of antineutrinos\footnote{The calculations presented in this paper are for antineutrinos only.  However, they are also valid for neutrinos emitted at beta-beam facilities.  In what follows, we refer to neutrinos as a generic term.} resulting from the decay of $\gamma=7$ and
$\gamma=14$ $^6$He ions.  The maximal energy of the neutrinos in
these cases are 55~MeV and 100~MeV, respectively.
We consider a
cylindrical
 detector with $r=4.5$~m (radius) and $h=15$~m (depth), placed 10~m from the end of the straight section, as done in Refs.~\cite{Balantekin:2005md,Balantekin:2006ga}.
\begin{table}[h]
\begin{center}
\begin{tabular}{ccccccc}
\hline $\gamma$ & $\left\langle E\right\rangle$ &
$\Gamma(E)$ & $E_{max}$ &
$\widetilde{\Psi}_{\max}$ & $N_{ev}$\\
\hline\hline 7 &  22.8 & 17.1 & 20.7 & 24.5 & 5782 \\
\hline 14 &  42.6 & 19.9 & 37.0 & 28.4 & 42393 \\
\hline
\end{tabular}
\caption{Average energy, energy dispersion 
$\Gamma(E)=(\left\langle E^{2}\right\rangle -\left\langle E\right\rangle ^{2})^{1/2}$ , peak-energy evaluated at $\widetilde{\Psi}_{\max}$ and peak-flux ($/10^{9}$)
at a cylindrical detector placed 10~m away from a low-energy beta-beam running $^6$He ions at $\gamma=7$ and $\gamma=14$.
$N_{ev}$ gives the number of events for one year (3 x $10^{7}$~s) for the
anti-neutrino scattering on protons from 
Eq.(\ref{eqn:dndt}) considering water as a target material.
Energies are in units of MeV, the flux in units of
m MeV$^{-1}$s$^{-1}$.}
\label{tbl:lowenergy}
\end{center}
\end{table}

\subsection{Off-axis neutrino fluxes at a standard beta-beam facility}
\label{sec:offaxis}
\begin{figure}[h]
\includegraphics[width=8cm]{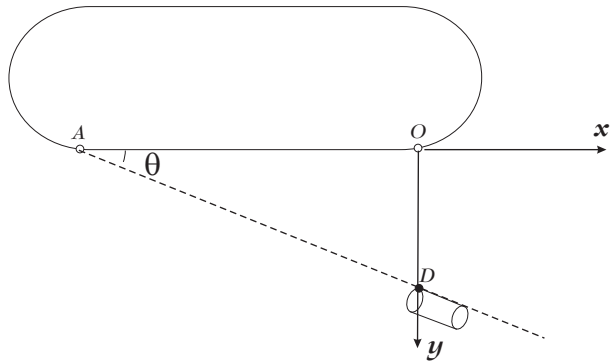}
\caption{The location of the single off-axis detector $D$ is shown
relative to the standard beta-beam storage ring with straight
section $AO$.} \label{fig:singledetector}
\end{figure}
Let us now consider the possibility of extracting low-energy
neutrinos from the standard beta-beam facility
\cite{Zucchelli:2002sa}, where ions are boosted at $\gamma=60-100$
and the neutrinos produced with energies up to\footnote{Note that the corresponding storage ring can not
be used to store ions with low $\gamma$, considered in
Sec.~\ref{sec:referencescenario}}
600~MeV.  The accelerated ions emit the
highest energy neutrinos along the boost direction. Therefore, by
placing the detector off the storage ring straight section
axis~(Figure~\ref{fig:singledetector}), one gets rid of the
highest energy component of the neutrino flux.  The idea of off-axis neutrino
 beams was first proposed in \cite{E889}. The highest energy
neutrinos reaching the point $D$ in Fig.~\ref{fig:singledetector}
will be emitted from the most distant point in the storage ring
straight section (point $A$).  If one wishes that only neutrinos
with energy less or equal to $E_{\rm cut}$ arrive at point $D$, the
angle $\theta =\angle ADO$ has to satisfy the condition
\begin{equation}
\label{eqn:theta}
\theta =\arccos \left[ \frac{\gamma -
(Q-m_{e})/E_{\rm cut}}{\sqrt{\gamma ^{2}-1}}\right]~,
\end{equation}
where $Q$ is the $Q$-value of the reaction.  The actual location of the
detector depends on the desired antineutrino cut-off energy.

\begin{figure}[h]
\includegraphics[width=8cm]{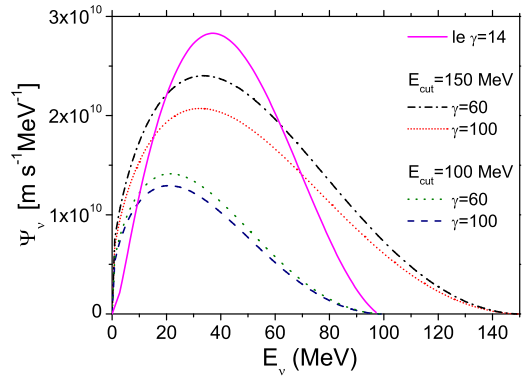}
\caption{ Off-axis antineutrino fluxes ($\times 20$) evaluated at point $D$ of Fig.~\ref{fig:singledetector}
for two different ion boosts and neutrino energy cuts.  The neutrino flux from a low-energy beta-beam (le)
at $\gamma =14$ is shown for comparison.}
\label{fig:pointdet}
\end{figure}
In Figure~\ref{fig:pointdet}, we compare the off-axis antineutrino
fluxes evaluated at point $D$, which lies on the perpendicular to
the storage ring straight section derived from the turning point
$O$ ($x=0$; see Fig.~\ref{fig:singledetector}).  The distance
$y={AD}= {AO}\cdot \tan\theta$ is determined using
~Eq.(\ref{eqn:theta}) and by constraining the maximum energy of
the neutrinos ($E_{\rm cut}$) reaching that point. The
presented results correspond to the cases where the $^{6}$He ions
are boosted at $\gamma =60$ and $\gamma=100$, and when $E_{\rm
cut}$ is set to 100 and 150~MeV.  In particular, if $\gamma =60$
 and $E_{\rm cut}=100$~MeV, the distance $y$ is 74.7~meters (Fig.~\ref{fig:singledetector}).
 If the ions are stored at $\gamma =100$, $y$ reduces to 61.4~meters, since the neutrino beam is more collimated.
 The main characteristics of such fluxes are summarized in Tables~\ref{tbl:pointdet1} and \ref{tbl:pointdet2}.
 In order to compare these results with the low-energy beta-beam fluxes of Table \ref{tbl:lowenergy},
 the values of $\widetilde{\Psi }_{\max }$ and $N_{ev}$ are normalized by the same detector volume.
\begin{table}[h]
\begin{center}
\begin{tabular}{cccccccc}
\hline $E_{\rm cut}$ & $\left\langle
E\right\rangle $ & $\Gamma(E)$ & $E_{max}$ 
& $\widetilde{\Psi}_{\max}$ & $N_{ev}$ & $y$ \\
\hline\hline
100 &32.7 & 19.6 & 21.0 & 0.64 & 626 & 61.4 \\
150 &  49.6 & 29.5 & 32.5 & 1.03 & 2998 & 48.0 \\
\hline
\end{tabular}
\caption{Same as Table~\ref{tbl:lowenergy} but for an off-axis flux at a point in space with coordinates $x=0$
and $y$ such that the maximum energy of the neutrinos is $E_{\rm cut}$ (value in the first column).
The ions are boosted at $\gamma=100$.  Energies are in units of MeV, the flux in units of
m MeV$^{-1}$s$^{-1}$.}
\label{tbl:pointdet1}
\end{center}
\end{table}
\begin{table}[h]
\begin{center}
\begin{tabular}{cccccccc}
\hline $E_{\rm cut}$ & $\left\langle
E\right\rangle $ &$\Gamma(E)$ & $E_{max}$
& $\widetilde{\Psi}_{\max}$ & $N_{ev}$ & $y$ \\
\hline\hline
100 &  32.9 & 19.7 & 21.4 & 0.70 & 693 & 74.7 \\
150 & 50.2 & 29.6 & 33.5 & 1.20 & 3512 & 56.0 \\
\hline
\end{tabular}
\caption{Same as Table~\ref{tbl:pointdet1} but for $\gamma=60$.}
\label{tbl:pointdet2}
\end{center}
\end{table}

From Tables~\ref{tbl:pointdet1} and~\ref{tbl:pointdet2}, one can see that the off-axis antineutrino flux
profiles are determined by the choice of $E_{\rm cut}$ (which determines the angle $\theta $) and come out
to be not very sensitive to the boost of the ions.  Note, however, that $N_{ev}$ is reduced by 10\% to 20\% when $\gamma$ changes from 100 to 60. The flux shapes are
 strongly asymmetric, centered at low energies, and have a long high-energy tail.

\begin{table}[h]
\begin{center}
\begin{tabular}{cccccccc}
\hline $E_{\rm cut}$ &  $\left\langle
E\right\rangle $ & $\Gamma(E)$ & $E_{max}$
& $\widetilde{\Psi}_{\max}$ & $N_{ev}$ & $y$ \\
\hline\hline
100 &  29.3 & 17.7 & 18.5 & 0.57 & 405 & 61.4 \\
150  & 43.6 & 26.3 & 28.0 & 0.87  & 1799 & 48.0\\
\hline
\end{tabular}
\caption{Same as Table~\ref{tbl:pointdet1} but for a cylindrical detector with $r=4.5$~m and $h=15$~m.}
\label{tbl:onedetector1}
\end{center}
\end{table}
\begin{table}[h]
\begin{center}
\begin{tabular}{cccccccc}
\hline $E_{\rm cut}$ &  $\left\langle
E\right\rangle $ &$\Gamma(E)$ & $E_{max}$ 
& $\widetilde{\Psi}_{\max}$ & $N_{ev}$ & $y$ \\
\hline\hline
100 & 30.3 & 18.2 & 19.5 & 0.63 & 497 & 74.7 \\
150 & 45.6 & 27.1 & 30.2 & 1.05 & 2405 & 56.0 \\
\hline
\end{tabular}
\caption{Same as Table~\ref{tbl:pointdet2} but for a cylindrical detector with $r=4.5$~m and $h=15$~m.}
\label{tbl:onedetector2}
\end{center}
\end{table}

Let us briefly discuss how the off-axis antineutrino flux changes
close to the point D of Fig.~\ref{fig:singledetector}. First, the
flux is clearly symmetric with respect to a rotation around the
straight section ${AO}$; it does not vary significantly along the
line ${AD}$ for distances of order $y<<{AO}$. Nevertheless, the
flux is very sensitive to variations of the angle $\theta$ and
reduces significantly once one moves away from the storage ring
straight section. Therefore, in order to have the highest count
rate at the detector, one should place it by aligning its
longitudinal part with ${AD}$. In Tables~\ref{tbl:onedetector1}
and \ref{tbl:onedetector2} we present the results for a
cylindrical detector with\footnote{This is the same detector
geometry as considered in the low-energy beta-beam scenario of
Section~\ref{sec:referencescenario}.} $r=4.5$~m and $h=15$~m.
 One can see that taking into account the physical size of the detector
reduces the flux at the peak only by about 10\%; however, it
strongly affects its high-energy tail and therefore the events
count rate.

From these results, it is clear that both the off-axis flux at the peak
intensity and the related number of events $N_{ev}$ are considerably
smaller -- by factors of 20-100 -- than those of the low-energy beta-beam
option.  Such drastic reduction clearly makes this option hardly realizable for low-energy neutrino physics applications, unless higher ion intensities are achieved.

\subsection{Alternative scenario: two off-axis detectors at a standard
beta-beam facility}
\label{sec:twodetectors}
In order to remove the high-energy neutrinos from the flux, the off-axis detector
should be placed relatively far away from the straight section
($y>50$~m).  This renders the intensities very low. It is worth
noting that the neutrino flux has almost the same energy
dependence at any point along the ${AD}$
(Figure~\ref{fig:singledetector}).  Furthermore, the flux
intensities along this
 line are inversely proportional to the distance from point $A$.  However, if one gets closer to point $A$,
 the signal start being contaminated by the high-energy neutrinos emitted from the bending part of the storage ring.

To overcome this difficulty, we introduce a novel technique which consists in comparing the
response of two off-axis detectors, placed in a specific configuration close to the storage ring axis.
 By using the subtraction procedure described below, one is able to extract the low-energy antineutrino
 flux and gain one order of magnitude in the intensity.

\begin{figure}[h]
\includegraphics[width=8cm]{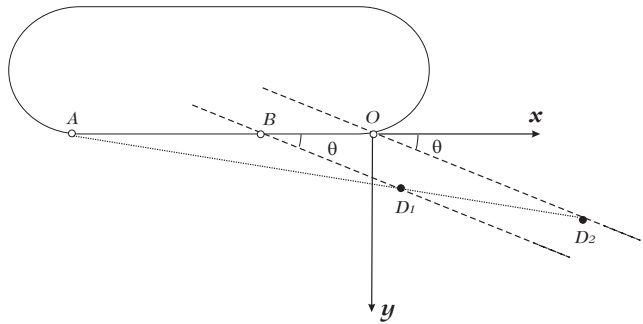}
\caption{Schematic view of the two off-axis detector locations, $D_1$ and $D_2$, relative to the straight
 section of the beta-beam storage ring, ${AO}$.}
\label{fig:twodetectorsfig}
\end{figure}
Let us consider the neutrino fluxes in two points $D_{1}$ and $D_{2}$, as shown in Figure~\ref{fig:twodetectorsfig}.
 We split the neutrino flux $\widetilde{\Psi}_{D_{1}}(E_{\nu })$ in point $D_{1}$ in two parts: one component produced in the
 segment $AB$ of the storage ring, i.e. $\widetilde{\Psi} _{D_{1}}^{(AB)}(E_{\nu })$, and the other produced in the segment
 $BO$, $\widetilde{\Psi} _{D_{1}}^{(BO)}(E_{\nu })$.  Since the triangles $ABD_{1}$ and $AOD_{2}$ are similar, the
 flux fraction $\widetilde{\Psi}_{D_{1}}^{(AB)}(E_{\nu })$ in point $D_1$ is proportional to the neutrino flux $\widetilde{\Psi}_{D_{2}}(E_{\nu })$
 in point $D_{2}:$
\begin{equation}
\label{eqn:fluxd1}
\widetilde{\Psi} _{D_{1}}^{(AB)}(E_{\nu })=\widetilde{\Psi} _{D_{2}}(E_{\nu })\frac{AO}{AB}
\end{equation}
The flux $\widetilde{\Psi}_{D_{1}}^{(BO)}(E_{\nu })$ can be obtained combining the responses of the two detectors located
in $D_{1}$ and $D_{2}$, by using the following subtraction procedure:
\begin{equation}
\label{eqn:fluxd1-d2}
\widetilde{\Psi}_{D_{1}}^{(BO)}(E_{\nu })=\widetilde{\Psi}_{D_{1}}(E_{\nu })-\widetilde{\Psi} _{D_{2}}(E_{\nu})\frac{AO}{AB}
\end{equation}
Note that this flux contains only neutrinos with energies less
than $E_{\rm cut}$, set by Eq.~(\ref{eqn:theta}). The subtracted
flux of Eq.~(\ref{eqn:fluxd1-d2}) has a similar energy dependence
as the flux at the point
\begin{eqnarray}
x &=&AD_{1}\cos \theta -AO~, \\
y &=&AD_{1}\sin \theta~,
\end{eqnarray}
but its intensity is higher by a factor of $y/y_{D_{1}}$. In
practice one will be subtracting the number of events measured by the
two detectors. Therefore the subtracted number of events
associated with the flux $\widetilde{\Psi}_{D_{1}}^{(BO)}(E_{\nu
})$ is obtained as follows :
\begin{equation}
\label{eqn:eventsd1-d2}
N _{D_{1}}^{(BO)}= n \int  dt \int \sigma (E_{\nu})
\widetilde{\Psi} _{D_{1}}^{(BO)}(E_{\nu }) dE_{\nu}=N _{D_{1}}-
N _{D_{2}}\frac{AO}{AB}
\end{equation}

In Section~\ref{sec:offaxis} we found that a $y$-distance of 48~m
to 75~m is required in order to get low-energy neutrinos from the
off-axis flux (Figure~\ref{fig:pointdet} and
Tables~\ref{tbl:pointdet1} and~\ref{tbl:pointdet2}). Here, the
detector $D_{1}$ can be placed very close to the storage ring.
This implies  a neutrino flux intensity enhancement by $\sim 10$.
The position of the detector at $D_{2}$ with respect to the
position of the detector at $D_{1}$ is fixed by the choice of the
desired maximal neutrino energy ($E_{\rm cut}$) of the subtracted
flux.  The same arguments are valid for the realistic,  large size
detectors: one should place two detectors having the same shape,
but the detector in $D_{2}$ should have  its linear dimensions
larger by a factor of $OD_{2}/BD_{1}$ than the detector in
$D_{1}$.

Once one considers a finite size detector, 
it is clear that the remote regions of the detector will see much
fewer neutrinos than the regions close to the subtraction axis
($BD_{1}$ for the detector located at $D_1$, or $AD_{2}$ for the
detector located at $D_2$).  For large size detectors their
shape and its orientation should play an important role. Probably
the best detector geometry would be the long and thin, hollow
inside, conus.  For such a geometry, one will have the
subtracted  fluxes very similar to the ones shown in
Figure~\ref{fig:pointdet}, but with an intensity higher by $\sim
y/y_{D_{1}}$. The cone-like detector geometry has been considered
in a recent theoretical study~{\cite{Amanik:2007zy}.  Nevertheless, the technical realization of such
detectors is expected to be difficult.

In order to show the sensitivity of the presented technique to the detector geometry,
we now consider the neutrino fluxes at four large detectors,
having standard shapes (spherical or cylindrical), whose
dimensions are given in Table~\ref{tbl:detectors}.  All four
detectors are taken to have the same volume; detector type d$_2$
also has the same shape as the reference detector of
Section~\ref{sec:referencescenario}. In Figure~\ref{fig:detcomp} 
we compare the subtracted neutrino
fluxes as well as differential number of events for the four detectors.
The latter are obtained by using the subtracted fluxes multiplied
by the anti-neutrino on proton cross sections
from Eq.(\ref{eqn:eventsd1-d2}) and considering that the detectors are
filled with water.  
The subtracted flux characteristics are given 
in Tables
\ref{tbl:twodetectors1} and \ref{tbl:twodetectors2}
for two different ion boosts, i.e. $\gamma=60,100$
and neutrino maximum energy cuts (100 and 150 MeV).
The cylindrical detectors are
considered to be placed longitudinally along the subtraction axis
($BD_{1}$ for the detector located at $D_1$ or $AD_{2}$ for the
detector located at $D_2$) as shown in
Figure~\ref{fig:twodetectorsfig}. In the case of spherical
detectors, the center of the first one is placed at $x=R_{\rm
det}$ and $y=R_{\rm det}+5$~m. For the disc detector, the upper
surface touches the subtraction axis and is inclined along it. One
can see that the longest detector (d$_{1}$) picks the most
neutrino flux: two times more than the spherical detector
(d$_{4}$).
\begin{figure}[h]
\includegraphics[width=7cm]{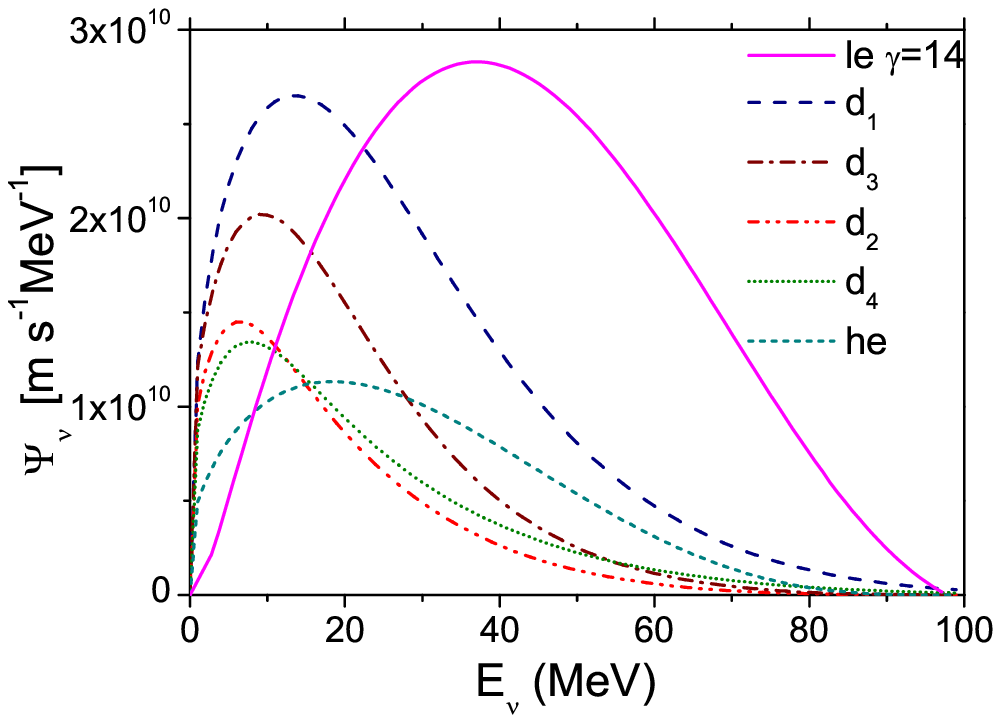}
\hskip 1cm
\includegraphics[width=7cm]{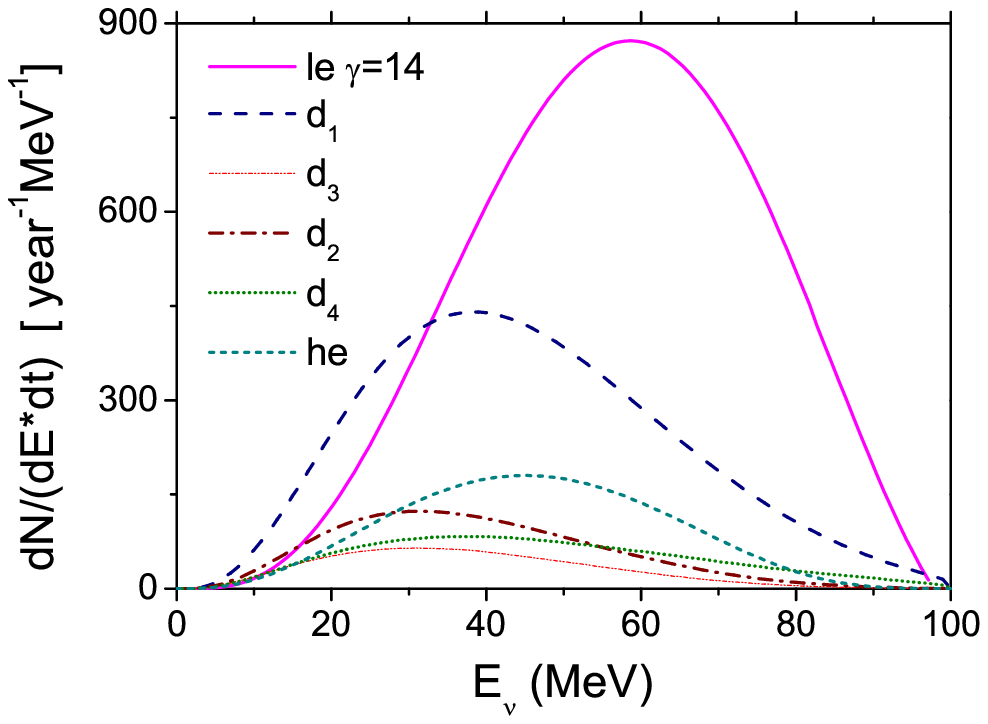}
\caption{Comparison of the different low-energy neutrino fluxes
(left panel) and the corresponding number of antineutrino-hydrogen
events for the water detectors (right panel). The resented results are
obtained for a standard beta-beam exploiting two detectors
off-axis (Figure~\ref{fig:twodetectorsfig}) and the subtraction
method described in the text (these fluxes are multiplied by 6).
The different curves correspond to four different detector
geometries, namely the cylinder-sausage ($d_1$), the
cylinder-normal ($d_2$), the cylinder-disc ($d_3$), and the
spherical (Table~\ref{tbl:detectors}). As a comparison, the fluxes
from a low-energy beta-beam (le), and for a single off-axis
detector $d_2$ as described in Section II.B (HE), 
are given. The last flux is multiplied by 20.}
\label{fig:detcomp}
\end{figure}
The flux profile is even more asymmetric than for the single
off-axis detector case (Figure~\ref{fig:pointdet}). Note that the
average energy is pushed towards much lower energies (around
10-20~MeV) compared to the low-energy beta-beam flux. The expected
intensities are significantly higher than in the case of the
single off-axis detector, but still a few times
weaker than for the low-energy beta-beam  discussed in
Section~\ref{sec:referencescenario}.
\begin{table}[h]
\begin{center}
\begin{tabular}{cccc}
\hline Name & Type & Parameters & Volume (m$^{3}$) \\ \hline\hline
d$_{1}$ & Cylinder-Sausage &
\begin{tabular}{l}
r~=~2.25 m \\
h~=~60.0 m
\end{tabular}
& 954 \\ \hline
d$_{2}$ & Cylinder-Normal &
\begin{tabular}{l}
r~=~4.50 m \\
h~=~15.0 m
\end{tabular}
& 954 \\ \hline
d$_{3}$ & Cylinder-Disc &
\begin{tabular}{l}
r~=~9.00 m \\
h~=~3.75 m
\end{tabular}
& 954 \\ \hline
d$_{4}$ & Spherical &
\begin{tabular}{l}
r~=~6.11 m
\end{tabular}
& 954 \\ \hline
\end{tabular}
\caption{Four different 954-ton detector geometries}
\label{tbl:detectors}
\end{center}
\end{table}

We have also studied the sensitivity to the ion boosts and $E_{cut}$ choices
as well as the $y$ distance from the storage ring. 
In Figure~\ref{fig:twosaussage}, we compare the subtracted
fluxes and differential number of 
events for type-d$_{1}$ detectors, when one is placed at
$y=5$~m from the storage ring straight section and its twin
detector is placed in such a way that the subtracted neutrino flux
is either $E_{\rm cut}=100$~MeV or 150~MeV.  The ions in the
storage ring are considered to be boosted to $\gamma =100$ and
$60$ (Tables \ref{tbl:twodetectors1} and \ref{tbl:twodetectors2}).
One can see that for large size detectors, a lower ion boost is
advantageous.  For example, one gains more than 30\% in intensity
by reducing the  boost from 100 to 60.
Figure~\ref{fig:twosaussagey} shows how the subtracted flux
intensities and  the differential numbers of
     neutrino events vary  by placing the detector at different distances
from the storage ring (the closest points are $y=$5, 7.5 and 10~m
away from the storage ring, respectively).  If the detector has a
small size compared to $y$, the subtracted intensity should scale
as $1/y$.   On the other hand, for the large detector we consider
one gains much less in intensity by placing it closer to the
straight section (e.g. $y=10$ and $y=5$~m fluxes differ only by
50\%).
\begin{table}[h]
\begin{center}
\begin{tabular}{ccccccccccccccccc}
\hline
& \hspace{0.5cm} & \multicolumn{6}{c}{$\gamma =100$} & \hspace{0.5cm} &
\multicolumn{7}{c}{$\gamma =60$} \\
& \hspace{0.5cm} & $\left\langle E\right\rangle $ &
 $\Gamma(E)$ & $E_{max}$ &
$\widetilde{\Psi }_{\max }$& $N^{tot}_{ev}$ &$N_{ev}$ &   $\sigma_{N_{ev}}$ 
\hspace{0.5cm} & $\left\langle E\right\rangle $ &
$\Gamma(E)$ & $E_{max}$ &
$\widetilde{\Psi }_{\max }$ & $N^{tot}_{ev}$   &  $N_{ev}$  &  $\sigma_{N_{ev}}$ \\
\hline\hline d$_{1}$ & \hspace{0.5cm} & 21.3 & 18.5 & 13.50 & 4.42
& 1.9(6) &  2487 & 1949 \hspace{0.5cm}
& 28.7 & 19.1 & 15.0 & 6.09 &  8.9 (5) & 4013  & 1334  \\
d$_{3}$ & \hspace{0.5cm} & 20.4 & 14.7 & 9.32 & 3.36 & 5.4(5) &879 & 1039
\hspace{0.5cm}
& 21.4 & 15.1 & 10.3 & 4.60 &  2.7(5) & 1363  & 735 \\
d$_{1}$ & \hspace{0.5cm} & 39.0 & 26.3 & 20.6 & 5.44
& 1.9(6) & 8162 & 1949 \hspace{0.5cm} & 41.27  & 26.8
& 23.3 & 7.63 &  8.9(5) & 13240 & 1334  \\ 
d$_{3}$ & \hspace{0.5cm} & 31.1 & 21.9 & 15.0 & 4.22 & 5.4(5) & 3481 & 1039
\hspace{0.5cm} & 33.07 & 22.5
& 17.2 & 5.93 & 2.7(5) & 5789 & 735 \\  \hline
\end{tabular}%
\end{center}
\caption{Same as Table~\ref{tbl:lowenergy} for the 
detector $d_1$ geometry of Table~\ref{tbl:detectors}. The upper (lower) line corresponds to
a cut-off energy of 
100~MeV (150~MeV).
The fluxes were
obtained with the subtraction method described in the text for a
standard beta-beam facility running $^6$He ions at $\gamma=100$
and $\gamma=60$. The quantity
$\widetilde{\Psi }_{\max }$ is multiplied by $10^{9}$.
For $N^{tot}_{ev}$, the number given in parenthesis corresponds to the
exponent. The statistical error $\sigma_{N_{ev}}$ associated to $N_{ev}$ is also given.}
\label{tbl:twodetectors1}
\end{table}
\begin{table}[h]
\begin{center}
\begin{tabular}{ccccccccccc}
\hline
& \hspace{0.5cm} & \multicolumn{3}{c}{$\gamma =100$} &
\hspace{0.5cm} & \multicolumn{3}{c}{$\gamma =60$} \\
& \hspace{0.5cm} & $\left\langle E\right\rangle $ &
 $\Gamma(E)$ & $E_{max}$ &
$\widetilde{\Psi }_{\max }$&  
\hspace{0.5cm} & $\left\langle E\right\rangle $ &
$\Gamma(E)$ & $E_{max}$ &
$\widetilde{\Psi }_{\max }$ \\ \hline \hline
d$_{2}$ & \hspace{0.5cm} & 18.5 & 14.3 & 6.50 & 2.42 &
\hspace{0.5cm}
& 25.3 & 14.6 & 7.29 & 3.31 \\
d$_{4}$ & \hspace{0.5cm} & 15.0 & 12.3 & 4.50 & 1.78 &
\hspace{0.5cm} & 24.0 & 20.6 & 5.43 & 2.42 \\
d$_{2}$ & \hspace{0.5cm} & 28.2 & 21.4 & 10.8 & 3.05 &
\hspace{0.5cm} & 29.95 & 22.0
& 12.7 & 4.31   \\
d$_{4}$ & \hspace{0.5cm} & 23.0 & 18.5 & 7.79 & 2.23 &
\hspace{0.5cm} & 24.8 & 19.2 & 9.30 & 3.16   \\ \hline
\end{tabular}
\caption{Sensitivity of the flux characteristics to the detector geometry
(Table \ref{tbl:detectors}).
The meaning of the presented values is the
same as in Table~\ref{tbl:twodetectors1}.}
\label{tbl:twodetectors2}
\end{center}
\end{table}
\begin{figure}[h]
\includegraphics[width=7cm]{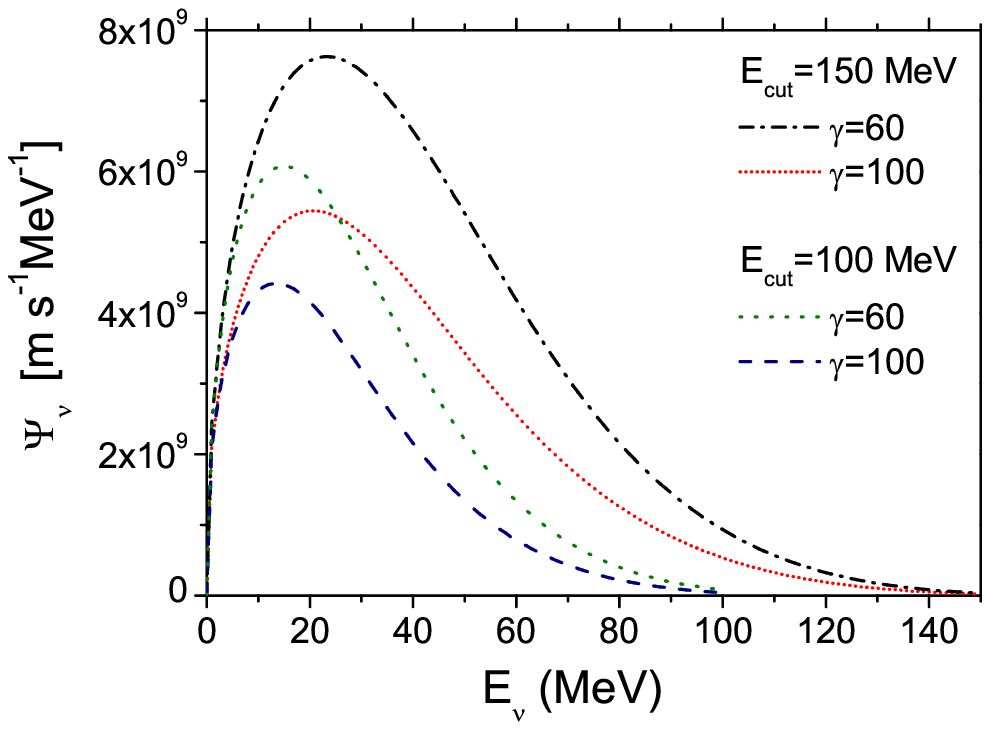}
\hskip 1cm
\includegraphics[width=7cm]{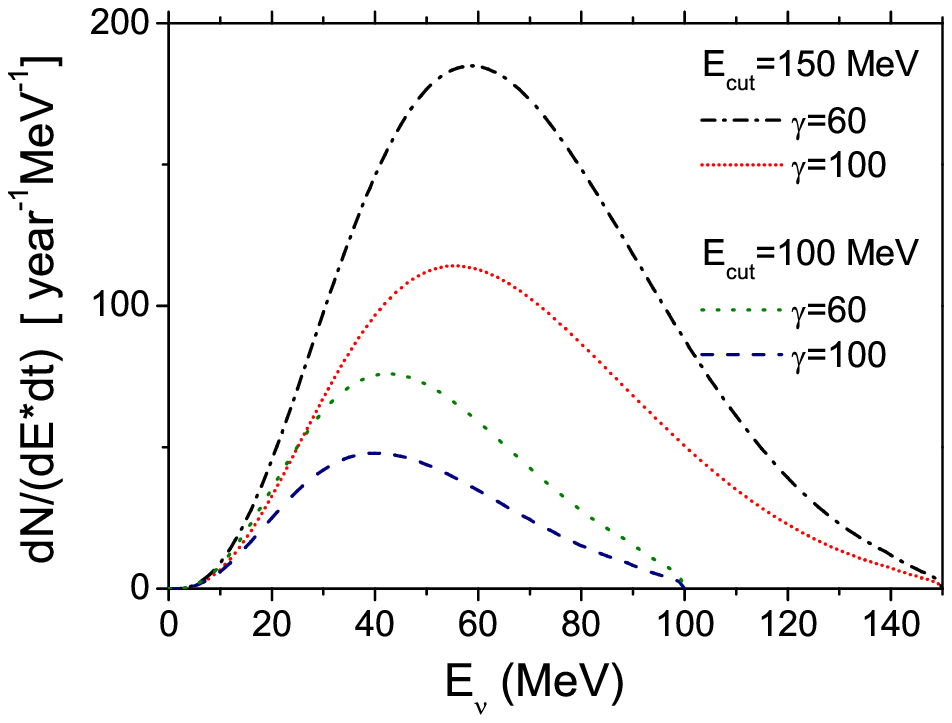}
\caption{Neutrino fluxes obtained by placing two detectors d$_1$
(Table \ref{tbl:detectors})  off-axis and using the subtraction
method described in the text. The different curves correspond to two ion boosts
and maximum neutrino energy cutoff. The first detector is located at $y=$5 m.}
\label{fig:twosaussage}
\end{figure}
\begin{figure}[h]
\includegraphics[width=7cm]{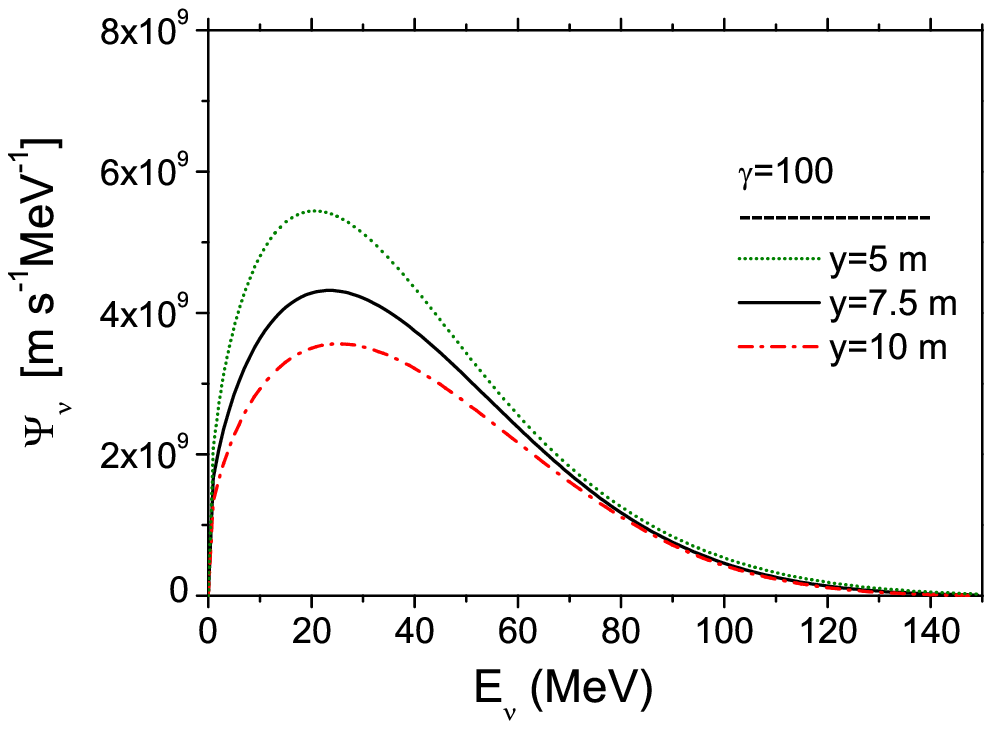}
\hskip 1cm
\includegraphics[width=7cm]{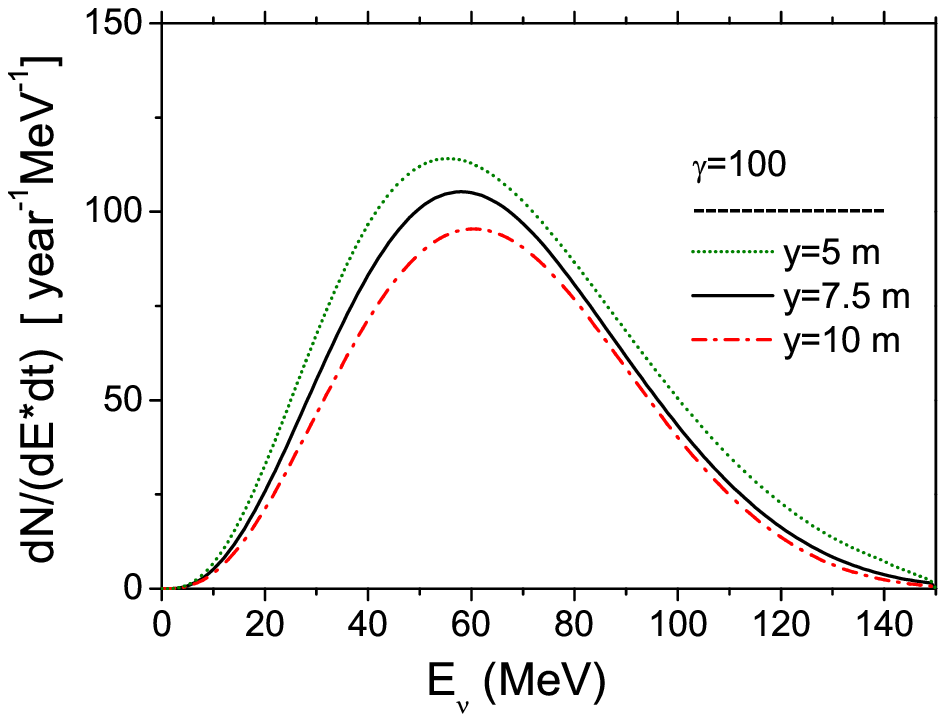}
\caption{Same as Figure \ref{fig:twosaussage} but for different $y$ locations.}
\label{fig:twosaussagey}
\end{figure}

For the $d_1$ and $d_3$ 
detector geometries, Table \ref{tbl:twodetectors1}  presents the total and subtracted
number of events associated to anti-neutrino proton scattering as well.
Since the cross sections grow
approximately as the square of the neutrino energy, the total
number of events in the detector is considerably larger than the
subtracted events at low energy. Therefore the statistical error
associated to the subtracted number of events is always
significant since it is determined by the error on the total
number of events: $\sigma_{N_{ev}}\approx\sqrt{2N_{ev}^{tot}}$. If
from the point of view of the characteristics of the subtracted
fluxes the low and high gamma value options are almost equivalent,
the low gamma option (and small radial size of the detector)
becomes crucial once the statistical error on the subtracted
number of events is considered. Indeed, for considered ion
intensities at the storage ring only the d$_1$  and d$_3$ 
(with $E_{cut}=150~$MeV)
detector scenarios
get statistical errors which are significantly small.
However, if further feasibility studies with improved production
methods show that a higher ion intensity can be achieved
the statistical errors for the $d_2,d_4$ geometries 
can become small. Besides, one should keep in mind that 
in the presented Tables we have shown results on protons
considering large volume detectors filled with water. Such a
choice is made only based on the fact that neutrino scattering on
protons is the only case for which reliable cross sections in a
wide energy range are available. The
use of heavy target nuclei should be more favorable for the
substraction technique, since then the detector volume can be
significantly reduced and render 
all the discussed geometries particularly
interesting.

As we were completing this paper, the authors of Ref.
\cite{Amanik:2007zy} presented an analysis of how neutrino
spectral shapes change at low-energy beta-beams depending on the
detector geometry and different locations within the same
detector. The analysis presented in the current paper and in Ref.
\cite{Amanik:2007zy} are complementary in exploring the potentials
of standard and low-energy beta-beam facilities, respectively.

\section{Conclusions}
\label{sec:conclusions}

In this article we explored the feasibility of extracting
low-energy neutrinos from the standard beta-beam facility by
placing detectors at off-axis. We found that with a single
off-axis detector the low-energy neutrino fluxes extracted are
rather small. We proposed  a two off-axis detector option,  which
allows, after suitable subtractions, a significant increase in the
number of low-energy events. The drawback of this method is an
increase in the statistical error.  The present work is based on
various preliminary assumptions such as the ion intensities.
Smaller statistical errors can then be achieved for the two
off-axis detector option if the ions circulate with higher
intensities in the storage ring. If that would be the case, the
option of a single detector would also have to be revisited since
the low-energy flux would be larger. In the case of the two
detectors option, the energy spectra of the neutrinos are pushed
to lower energies than for the low-energy beta-beam. The covered
energy range is of interest for fundamental tests and for
core-collapse supernovae physics. We also studied the dependence
of the flux intensity and energy spectrum on the location and
geometry of the detectors.

We conclude that the option of two off-axis detectors at a standard beta-beam facility
might be an alternative to the reference scenario
of a low-energy beta-beam facility for the realization of low-energy neutrino experiments.

\begin{acknowledgments}
The authors acknowledge the CNRS-Etats Units 2005 and 2006 grants which have
been used during the completion of this work.  This work was also supported in part by the U.S. National Science
Foundation Grant No. PHY-0555231 at the University of Wisconsin, and in part
by the University of Wisconsin Research Committee with funds granted by the
Wisconsin Alumni Research Foundation.  C.V. and R.L. acknowledge the financial
support of the EC under the FP6 "Research Infrastructure Action-Structuring the European Research Area" EURISOL DS Project; Contract No. 515768 RIDS.
\end{acknowledgments}

\end{document}